\documentclass{emulateapj}
\usepackage{graphicx}

\newcommand{\lya}{Ly$\alpha$~}
\renewcommand{\thefootnote}{}
\def\see{\mbox{$^{\prime\prime}$}}
\shorttitle {The earliest galaxies and the epoch of reionization }
\shortauthors{Pentericci et al.}

\begin{document}
\title{Spectroscopic confirmation of z$\sim$ 7 LBGs:
probing the earliest galaxies and the epoch of reionization}

\author{L.Pentericci\altaffilmark{1}
A. Fontana \altaffilmark{1} 
E. Vanzella\altaffilmark{2}
M. Castellano \altaffilmark{1} 
A. Grazian \altaffilmark{1}
M. Dijkstra \altaffilmark{3}
K. Boutsia \altaffilmark{1} 
S. Cristiani \altaffilmark{2} 
M. Dickinson\altaffilmark{4}
E. Giallongo \altaffilmark{1}
M. Giavalisco \altaffilmark{5}
R. Maiolino \altaffilmark{1} 
A. Moorwood\altaffilmark{6*}
P. Santini \altaffilmark{1}}

\affil{$^1$ INAF Osservatorio Astronomico di Roma,  Via Frascati 33,00040 
Monteporzio (RM), Italy}

\affil{$^2$ INAF Osservatorio Astronomico di Trieste, Via G.B.Tiepolo 11, 
34131 Trieste, Italy}

\affil{$^3$ Max-Planck-Instit für Astrophysik, Karl-Schwarzschild-Str. 1, 85741 Garching, Germany}

\affil{$^4$ National Optical Astronomy Observatory, PO Box 26732, Tucson, AZ 85726, USA}
\affil{$^5$ Department of Astronomy, University of Massachusetts, 710 North Pleasant Street, Amherst, MA 01003 }

\affil{$^6$ European Southern Observatory, Karl-Schwarzschild Strasse, 85748 Garching bei Munchen, German}
\email{laura.pentericci@oa-roma.inaf.it}

\footnote{* We would like to dedicate this paper in memory of Alan Moorwood, who left us a few days before the paper was submitted. Alan was fundamental to the development of Hawk-I, which enabled this survey and many other important observing programs.  He had clear foresight of the instrument's impact on the search for the highest redshift galaxies. More importantly, he always urged us to obtain spectroscopic confirmation of the candidates, and he was eagerly awaiting the results of this effort.}
\begin{abstract}
 We present  the final results from our ultra-deep spectroscopic campaign with FORS2 at the ESO/VLT for the confirmation of $z\simeq7$ ``z--band dropout'' 
candidates 
  selected from our VLT/Hawk-I imaging survey over three independent fields. 
In particular we report on  two newly discovered galaxies at redshift $\sim$
6.7 in the NTT deep field: both galaxies show 
a \lya emission line with rest-frame EWs of the order 15-20\AA\ 
and luminosities of 2-4$\times 10^{42} erg s^{-1}$.  
We also present the results of ultra-deep observations of a sample of i-dropout galaxies,
 from which we set a solid upper limit on the fraction of interlopers. 

Out of the 20 z-dropouts observed we confirm 5 galaxies at $6.7 < z < 7.1$. 
This is systematically  below the expectations 
drawn on the basis of lower redshift observations: in particular 
there is a significant lack of objects with intermediate \lya EWs (between 20 and 55 \AA).
We conclude that the trend for
the fraction of  \lya emission in LBGs  that is constantly 
increasing  from z$\sim$3 to z$\sim$6 is most probably 
reversed from $z\sim 6$ to  z$\sim$7.
 Explaining the observed rapid change in the LAE fraction among the drop-out population with reionization requires a fast evolution of the neutral fraction of hydrogen in the Universe. Assuming that the Universe is completely ionized at z=6 and adopting  the semi-analytical models of Dijkstra et al. (2011), we find that our data require a change of the neutral hydrogen fraction of the order $\Delta\chi_{HI} \sim 0.6$ in a time $\Delta z\sim1$, 
provided that the escape fraction does not increase dramatically over the same 
redshift interval.

\end{abstract}

\section{Introduction}
The epoch of reionization marks a major phase transition of the Universe,
during which the intergalactic space became transparent to UV photons. 
Determining when this occurred and the physical processes 
involved represents the latest frontier in observational cosmology. 
Over the last few years, searches have intensified 
to identify the population of high-redshift galaxies that might be responsible 
for this process. 
Recent data  from the Hubble Space Telescope WFC3 infrared camera 
provided a dramatic advance accessing  the faint side of the z$\sim 7 $ UV Luminosity Function (LF) 
\citep{McLure2010,Bouwens2010a,Grazian2011},
while wide field ground-based surveys are starting to detect the brightest 
galaxies at z$\sim 7$ \citep{Ouchi2009,Castellano2010a,Castellano2010b}. 
These surveys provided evidence that the UV LF  
evolves significantly with redshift \citep{Wilkins2010}, that the UV 
continuum of galaxies appears progressively bluer with increasing z 
(\cite{Fin2010,Bouwens2010b} but see also \cite{Dunlop2011}) and that their stellar masses are on average smaller than those of their lower-z counterparts
\citep{Labbe2010}, while the morphologies remain essentially unchanged from z
$\sim$6 to z$\sim$7 \citep{Oesch2010}.

Unfortunately, these results are entirely based on color selected samples.  The
lack of spectroscopic redshifts prevents us from deriving firm conclusions
based on these early samples.  First, an unknown fraction of interlopers
biases the estimates of size and redshift distribution which are needed to
compute the LF. The lack of spectroscopic redshift also introduces
degeneracies in the SED fitting that make the masses and other physical
properties poorly determined.  

Finally only spectroscopic observations of
high-z Ly$\alpha$ emission have the potential to provide a definitive answer to
the major question of how and when reionization occurred \citep{Dijkstra2011,Dayal2010}.
The \lya\ emission  line is a powerful diagnostic since  it is
easily erased by neutral gas outside galaxies. Its observed strength in
distant galaxies is therefore a sensitive probe  of the latest time when
reionization was completed (e.g. Robertson et al. 2011). 
Indeed some recent studies pointed to a
decrease in the number of \lya\ emitters (LAEs) at z$\sim$7 \citep{Ota2010} and 
failed to  detect 
bright LAEs at even higher redshift \citep{Hibon2010}. In a recent work 
we also found a significant lack of intense \lya\ in ultradeep spectra of candidate  $z\sim 7$ Lyman break galaxies 
\citep{Fontana2010} (F10 from here on). All these works seem to indicate that we might be approaching the epoch when the intergalactic medium was  
increasingly neutral.

To address these issues we have carried out a systematic follow-up of our
sample of z$\sim 7$ candidates: galaxies were initially selected as z-dropouts in three independent fields, the GOODS-South field (Giavalisco et al. 2004), the New Technology Telescope Deep Field (NTTDF, Arnouts et al. 1999, Fontana et al. 2000) and the BDF-4 field 
(Lenhert \& Bremer 2003) from deep near-IR  Hawk-I observations, covering a total area of 200 arcmin$^2$, complemented by  WFC3 observations
 \citep{Castellano2010a,Castellano2010b}. We then carried out deep spectroscopic observations with FORS2.
The first results were published in F10 
where we reported the confirmation of a weak  Ly$\alpha$ emission line at $z=6.97$ in the GOODS-South field, and in Vanzella et al. 2011 (V11) were we presented  two galaxies
in the BDF-4 field, showing  bright Ly$\alpha$ emission at  $z \approx 7.1$.

In this final paper we present the results on the z-dropouts obtained 
in the NTT Deep Field, where we confirm two further galaxies
at z$\geq 6.7$ through the presence of \lya\ emission. 
We also present and briefly discuss the  results obtained 
on the i-dropout  galaxies that were observed in the three separate fields
as secondary targets.

Finally we give an overview of the results obtained on the entire z-dropout 
sample and  compare them  to the expectations based on lower redshift 
observations
to determine the redshift evolution of the fraction of \lya\ emitters in Lyman break galaxy samples. We discuss the implications of our results 
on the epoch of reionization.

All magnitudes are in the AB system, and we adopt
$H_0=70$~km/s/Mpc, $\Omega_M=0.3$ and $\Omega_{\Lambda}=0.7$.

\section{NTTDF target selection and Observations}

The targets were selected as candidate z$\sim 7 $ galaxies 
according to the criteria described
extensively in C10b: besides the  five  $z$--band dropout
candidates listed in Table 3 of that paper, two further  $z$--band dropouts
were also included in the slits. These two objects (NTTDF-474 and NTTDF--2916) 
have a Y--band magnitude that is slightly fainter than the completeness limit  applied in C10b to derive the luminosity function (they have respectively Y=26.53 and Y=26.67)
but for the rest their photometry is  entirely consistent with the z-dropouts selection.
In the remaining slits we placed  several i-dropout candidates (see Sec 3.2)
  and other interesting objects such as few of the 
interlopers described in C10a which will be discussed in a future paper. 

Observations were taken in service mode with the FORS2 spectrograph on
the ESO Very Large Telescope, during July-August 2010. 
We used the 600Z holographic grating, that provides the highest
sensitivity in the range $8000-10000$\AA\  with a spectral resolution
 $R\simeq 1390$ and a sampling of 1.6\AA\  per pixel for a 1\see~slit. The
data presented here come from the co-addition of 86 spectra of 665
seconds of integration each, on a single mask, for a total of
57120 s (15.9 hr), with median seeing around 0.8\see. 
The sources have been observed through slitlets 1\see~wide by 12\see~long.
Series of spectra were taken at two different positions, offset
by $4"$ (16 pixels) in the direction perpendicular to the dispersion.

Standard flat-fielding, bias subtraction and wavelength calibration 
have been applied as in \cite{Vanzella2009,Vanzella2011} and F10.
The sky background has been subtracted between 
consecutive exposures, exploiting the fact that the target
spectrum is offset due to dithering.
Before combining frames, 
particular care has been devoted to the possible offset along the wavelength 
direction, by measuring the centroids of the sky lines in the wavelength 
interval 9400-9900\AA. We have also carried out the sky subtraction 
by fitting a polynomial function to the background.
The two approaches provide consistent results.

Finally, spectra were flux-calibrated using the observations of
spectrophotometric standards. 
Based on the analysis of the standard 
star  observed with the same setup of science targets, we derive 
that  the relative error due to flux calibration is within
the 10\%.
We expect slit losses to be small, given the
extremely compact size of the targets and the excellent seeing during the observations and therefore we neglect them 
in the subsequent discussion.

We remark that we used exactly the same set up and integration times as
in the previous observations, presented in F10 and V11. The reduction
 and calibration procedures employed were also the same: 
this implies that we have a very homogeneous set of observations
and the spectra have a uniform final depth.

\section{Results}
\subsection{Observations of z-dropouts}
In Table 1 we present the results for all the z-dropouts observed in the 
NTT deep field. The galaxies observed in the GOODS field were already discussed 
in F10. We also discuss the z-dropouts in the BDF-4 field that were observed but not spectroscopically confirmed (the two \lya\ emitters at $z > 7$ in this field were extensively described in V11).
We discuss all sources where some feature is detected and 
the possible  identifications(s).
\begin{table*}
\scriptsize
\begin{center}
\caption{Spectroscopic properties of observed z-dropouts in the NTTDF \& BDF-4 fields \label{tbl-2}}
\begin{tabular}{llllllll}
ID         & RA       & Dec  & Y & feature            &  identification  & flux  & S/N 
\\
           &          &      &   &                    &                  & $erg s^{-1} cm^{-2}$ \\
\\
\hline
NTTDF--474  &181.373976 & -7.664256 & 26.50 & line@9270 \AA     &  Ly$\alpha$@z=6.623 & 3.2$\times 10^{-18}$ & 7  
\\ 
\\
NTTDF--1479 &181.342892 & -7.681269 & 26.12 & no feature   &      & 
\\
\\
NTTDF--1632 &181.385721 & -7.683538 & 26.44 & no feature        & 
\\
\\
NTTDF--1917 &181.321143 & -7.687666 & 26.32 & line@8415 \AA     & Ly$\alpha$/H$\beta$     & 3.2$\times 10^{-18}$ & 9  
\\
\\
NTTDF--2916 &181.308925 & -7.702437 & 26.64 & no feature        &
\\
\\
NTTDF--6345 &181.403901 & -7.756190 & 25.46 & line@9364.5  \AA  &  Ly$\alpha$@z=6.701 & 7.2$\times 10^{-18}$ & 11  
\\
\\
NTTDF--6543 &181.383383 & -7.759527 & 25.75 & line@9566(tentative) &  associated with another object &\\
\\
BDF4--2687   &337.045197 & -35.155960  & 26.15 &  no feature   \\ 
\\
BDF4--2883   &337.028137 & -35.160046  & 26.15 & no feature  \\
\\
BDF4--5583   &337.047363 & -35.204590  & 26.65 & no feature \\
\\
BDF4--5665   &337.051147 & -35.205826  & 26.64 & no feature    \\ 

\end{tabular}
\end{center}
\end{table*}

{\bf NTTDF--474}
This object shows a faint line at 9270 \AA\ and no continuum, in agreement 
with the faint broad band magnitude. 
We exclude an identification with one 
of the [OII] components  since we would see the other 
component as well (the line falls on an area that is free from sky lines) and 
 given our resolution (see also the discussion in V11).
We also exclude an identification with one of the two [OIII] components or H$\beta$: 
in these case we would simultaneously detect the 3 lines (at 5007, 4959 
and 4861\AA\ respectively) which all fall in sky-free portions of the spectrum.
Finally, if the line were H$\alpha$ from a low redshift object, then 
the large drop observed between the z and Y-band of more than 1.6 magnitude,  
as well as the non-detections  in the V, R and I band down to AB=29, would
 be not consistent with the redshift.
\\
We therefore conclude that the most likely identification is $Ly\alpha$ at z=6.623.
This redshift is in excellent agreement with the photo-z =6.8.
The final extracted 1-dimensional and 2-dimensional spectra of this object are presented in Fig. 1.
Note that the line shows no clear asymmetry, but given the low S/N 
we cannot make any firm conclusions based on this.
\\
 {\bf NTTDF--1917}
This objects shows an emission line at 8415 \AA\ with 
flux 3.2$\times 10^{-18}$ erg s$^{-1}$ cm$^{-2}$. 
In the same slit we observe another emission line at  8640 \AA: this line is much more diffuse (spatially) and 
has an offset along the spatial axis of $\sim 2-3$ pixels corresponding to 0.7$''$. The  offset 
is consistent  with the position of a close-by diffuse galaxy.  
We believe that this second line is associated to the nearby object, whose light 
enters the slit. 
\\
The line at  8415 \AA\ cannot be identified with [OII] because we do not  
observe the typical double component. 
An identification with 
[OIII] at z=0.68 (either of the components) is also quite unlikely since 
the other [OIII] component and H$\beta$ should be visible in the spectra.
In case of identification with H$\beta$ at z=0.73,
 the [OIII] component 5007 would fall on the top of a skyline so we could possibly 
fail to observe it.
Finally for H$\alpha$ at $z=0.28$, the implied rest-frame EW would
 be unusually high ($> 300\AA$): the shape of the SED also disfavors this 
option, given the non-detections in the V, R and I band.

However even the identification of the line as a Ly$\alpha$ emission poses
some problems: the implied redshift of $z=5.92$ is hardly compatible with the
z-dropout selection function  (see the redshift distribution 
in Castellano et al. 2010b). Indeed 
the  best-fit zphot solution is  at 6.87, mainly based 
on the large color term $z-Y$. The fit is however still acceptable
(reduced chi squared 1.4) also for the solution at $z=5.92$,
 although the object would have to be a very young, dusty galaxy  with an unusually high star formation rate.

For the rest of the paper, we will consider this object as an interloper of the z$\sim 7$ sample.
\\
 {\bf NTTDF--6345}
This objects shows a line at 9364 \AA\ with flux 
  7.7$\times 10^{-18}$ erg s$^{-1}$cm$^{-2}$. The flux estimate is actually somewhat 
uncertain due to two problems: the line falls quite close to a bright
 skyline  (so the contribution of the red wing might have been 
poorly estimated); there is a bright object placed  4$''$
 away from our galaxy that  falls in the slit. The distance is exactly the 
same as our dither patter: as a result the zero flux level has some uncertainty.
For the identification of the emission line,
 we follow the same consideration as for NTT-474 to exclude the lower redshift interpretation.
Note that despite the vicinity of the skyline, we exclude the possibility 
that the emission line might be [OII] 
because the second component should be 
clearly visible 2 pixels outside the skyline.
\\
We conclude that the  line is Ly$\alpha$ emission at redshift  6.701,
 in excellent agreement with the photometric redshift of 6.73.The extracted 1-dimensional and the 2-dimensional spectra of this object are presented in Fig. 1.
\\
 {\bf NTTDF--6543}
In the spectrum of this object, we tentatively detect an  emission 
line falling exactly on the top of a skyline
at $\lambda$9566\AA.  However the emission is extremely faint, 
at the very limit of our detection 
sensitivity:  in addition, there is a significant offset between its  
spatial position  and the expected position of our  target along the slit 
(approximately 3 pixels, 0.75$''$).
Given the high uncertainty both in the reality of the line and in its association to 
the  z-dropout candidate, we will consider this object as unconfirmed. 

\begin{table*}
\scriptsize
\begin{center}
\caption{Spectroscopic properties of observed i-dropouts\label{tbl-2}}
\begin{tabular}{llllllllll}
ID            & RA    & Dec  &$z_{AB}$ & feature             &  ident.  & redshift        & Flux &  EW &   Quality \\
              &       &      &         &                     &         &                 &$erg s^{-1} cm^{-2}$ & \AA & \\
\hline
NTT--1806     & 181.365212 & -7.685984 & 26.81 & doublet@8709  & [OII]             & 1.335  &       &      & B \\
NTT--4025     & 181.331168 & -7.718568 & 26.51 & linee@8072    & Ly$\alpha$      & 5.638  & 1.6$\times 10^{-18}$   & 7    & B  \\
NTT--2313     & 181.380362 & -7.693486 & 26.16 & continuum +break at 8550 &      & 6.07   & --    &      & B  \\
NTT--7173     & 181.351769 & -7.767678 & 26.73 & line@8474     &  Ly$\alpha$     & 5.969  & 3.2   & 18  & B  \\
NTT--7246     & 181.380139 & -7.770059 & 26.02 & line@8175+continuum&Ly$\alpha$  & 5.724  & 4.24  & 12 & A  \\
BDF--2203     & 336.957961 & -35.14716 & 26.39 & line@8656+continuum&Ly$\alpha$  & 6.118  & 2.5   & 9.9 & A  \\
BDF--3367     & 336.956062 & -35.167719& 25.87 & continuum+break &               & 5.73   & --    &      & B  \\
BDF--4085     & 336.957328 & -35.181049& 26.13 & line@8750+continuum&Ly$\alpha$  & 6.196  & 23.5  & 110   & A  \\
BDF--4568     & 336.960484 & -35.188689& 26.46 & faint diffuse continuum &       & low z  & --    &      & C  \\
BDF--5870     & 336.955504 & -35.208964& 26.67 & line@8064+continuum &Ly$\alpha$ & 5.632  & 1.8   & 10   & A  \\
BDF--3995     & 336.979035 & -35.179460& 26.28 & line@8752+continuum &Ly$\alpha$ & 6.198  & 2.1   & 7.3  & A  \\
BDF--2890     & 336.991181 & -35.160187& 25.51 & continuum+break at 8150&        & 5.70   & --    &      & C  \\
BDF--5889     & 336.998416 & -35.209229& 26.59 & line@7953 \AA       &Ly$\alpha$ & 5.540  & 2.0   & 9.8  & B  \\
GOODSS--15052 & 53.145019  & -27.762773	 & 26.58 & line@8442\AA   & Ly$\alpha$   & 5.942  & 3.0   & 14   & B  \\
GOODSS--79    & 53.122540  & -27.7605000 & 26.56 & line@8424\AA   &  Ly$\alpha$  & 5.928  &3.65   & 16   & A  \\
GOODSS--12636 & 53.151438  & -27.720971  & 26.70 & line@8426\AA   &  Ly$\alpha$  & 5.929  &7.15   & 45   & A   \\
GOODSS--48    & 53.159500  & -27.7714443 & 26.47 & no feature     &              &        &--          & \\
\hline
\end{tabular}
\end{center}
\end{table*}

\subsection{Observations of i-dropouts}
As mentioned in Section 2, some of the slits which were not filled 
with primary z-dropouts candidates 
were used to observe candidate z$\sim$ 6 galaxies: this was done also 
for the previous observations  of the GOODS south and BDF fields,
 and we report all results here.
The i-dropouts were selected according to the following color criteria:
for the GOODS-S field the selection is the same as in V10, namely
$(i_{775}-z_{850})>1.3$ and $S/N(B_{435})<2$ and $S/N(V_{606})<2$;
for the BDF and NTT fields the selection was $I-Z> 1.3$ and  $S/N< 2$ in 
all band bluewards of the I-band. 
We remind that our initial catalog was  Y-selected, therefore all objects were initially detected in the Y-band with a limit of  to AB=27 (C10a). 
\\
In Table 2 we report the results for all i-dropouts that were observed 
in the three independent fields. We report the line fluxes, total z-band magnitude and the 
equivalent width (EW) of the lines, after taking out the contribution of the  line itself
to the broad band magnitude for $z>5.8$. The quality flags (A, B and C) are set
accordingly to V09.
We remark that these are amongst the deepest observations ever made 
on z$\sim$6 galaxies, since typically all other studies employ much shorter integration times 
(4-8 hours see e.g. V10, Stanway et al. 2009).
In total, 17 i-dropouts were observed: we determine an unambiguous  
redshift for 15 of the  candidates,  of which 14 are at redshift between 5.5 and 6.2 
and one is an [OII] emitters at z=1.335. 
Of the remaining 2, one has no feature detected and in one  we see faint continuum over most of the spectral range 
but we observe no break: moreover this continuum  is extremely extended spatially (more than 15 pixels),
  hence its identification as a high redshift object is highly dubious. 

The immediate conclusion that we can draw from these observations is that the i-dropout selection is 
indeed very robust and that most of the objects that usually remain undetected 
in the medium-deep spectroscopic follow up, are indeed at redshift $\sim 6$ .
We can set a very robust upper limit for the interloper fraction as $< 18\% (3/17)$ 
(in the worst case where the uncertain object and the undetected one  are both 
interlopers).
We also note that the fraction of galaxies with \lya\ in emission 
is actually quite high: 
11 candidates show a \lya\ line in emission, although often with small EWs.

\section{Redshift evolution of the Ly$\alpha$ fraction }
\subsection{Total sample}
In our  spectroscopic campaign, we have observed a large number of 
z-dropout candidates: in total, 7  were observed in the GOODS-S field (F10) 
of which one was confirmed, 
6 were observed in the BDF of which two were confirmed (V11) 
and finally 7 were observed in the NTTDF of which two were confirmed (this paper).
In Table 3 we report the basic  properties of the 5 confirmed galaxies  
at $z > 6.5$, and in  Figure 1 we show their 1D and 2D spectra.
Only two of the confirmed  galaxies have relatively 
bright \lya emission 
and would be also selected as Lyman alpha emitters, 
given their $EW>20$\AA.
Together with the galaxy at z=6.96, initially 
 found by \cite{Iye2006} as a Lyman $\alpha$ emitter but with continuum
colors that fulfill the Lyman break selection \citep{Ouchi2009},
our galaxies form at the moment of writing the only set of spectroscopically
  confirmed z$\sim 7$ LBGs.
\begin{figure*}
\epsscale{1.0}
\plotone{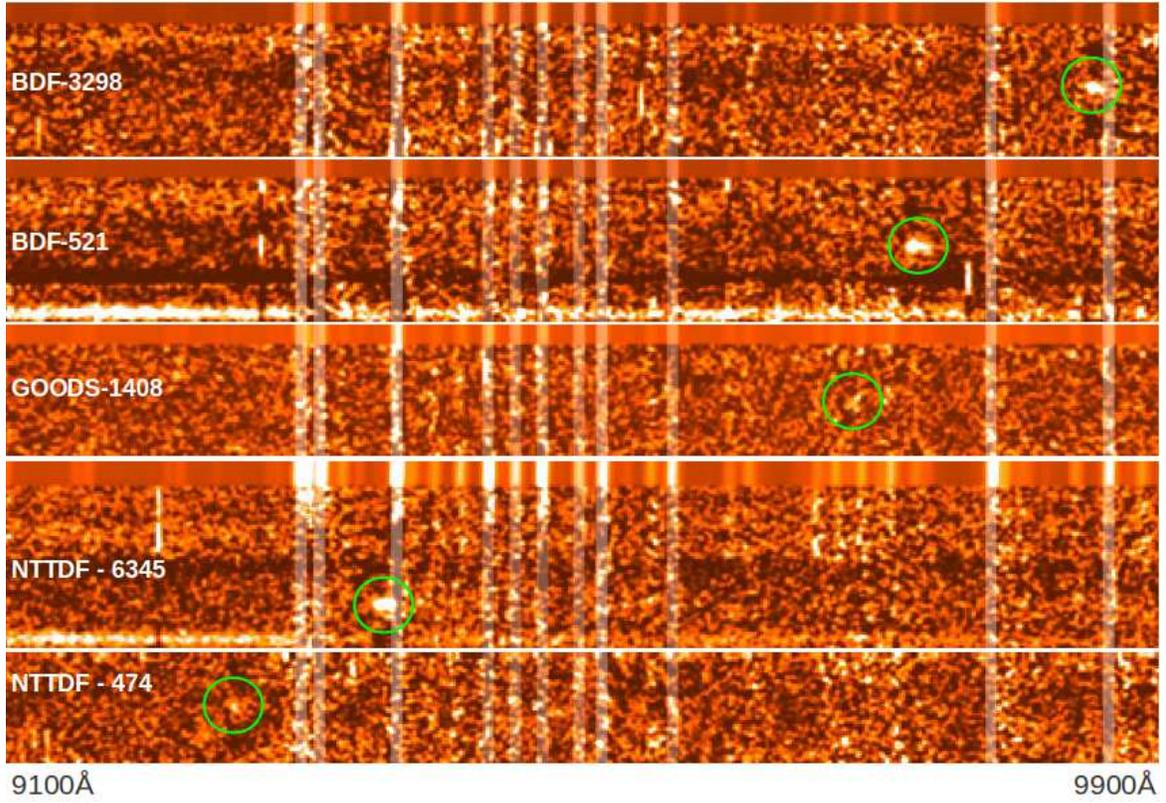}
\label{spectra2d}
\epsscale{0.8}
\plotone{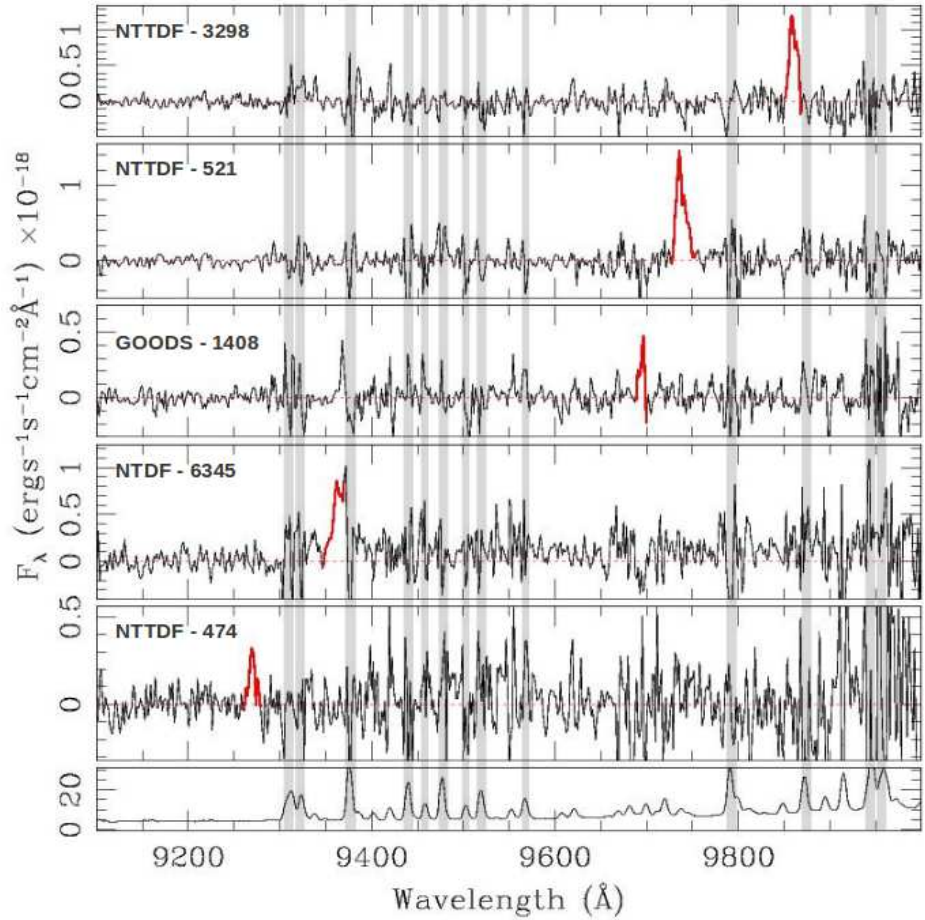}
\caption{Top panel: the 2-dimensional spectra of the 5 confirmed galaxies.
The spectra have been  filtered 
with a  the FILTER/ADAPTIV  Midas task using a box of 3x3 pixels. 
Bottom panel: the extracted 1-D spectra of the 5 confirmed 
galaxies and of the sky.
The bands indicate the positions of the strongest skylines that produce 
residuals in the reduced spectra.}
\label{spectra1d}
\end{figure*}

\begin{figure}
\epsscale{1.0}
\plotone{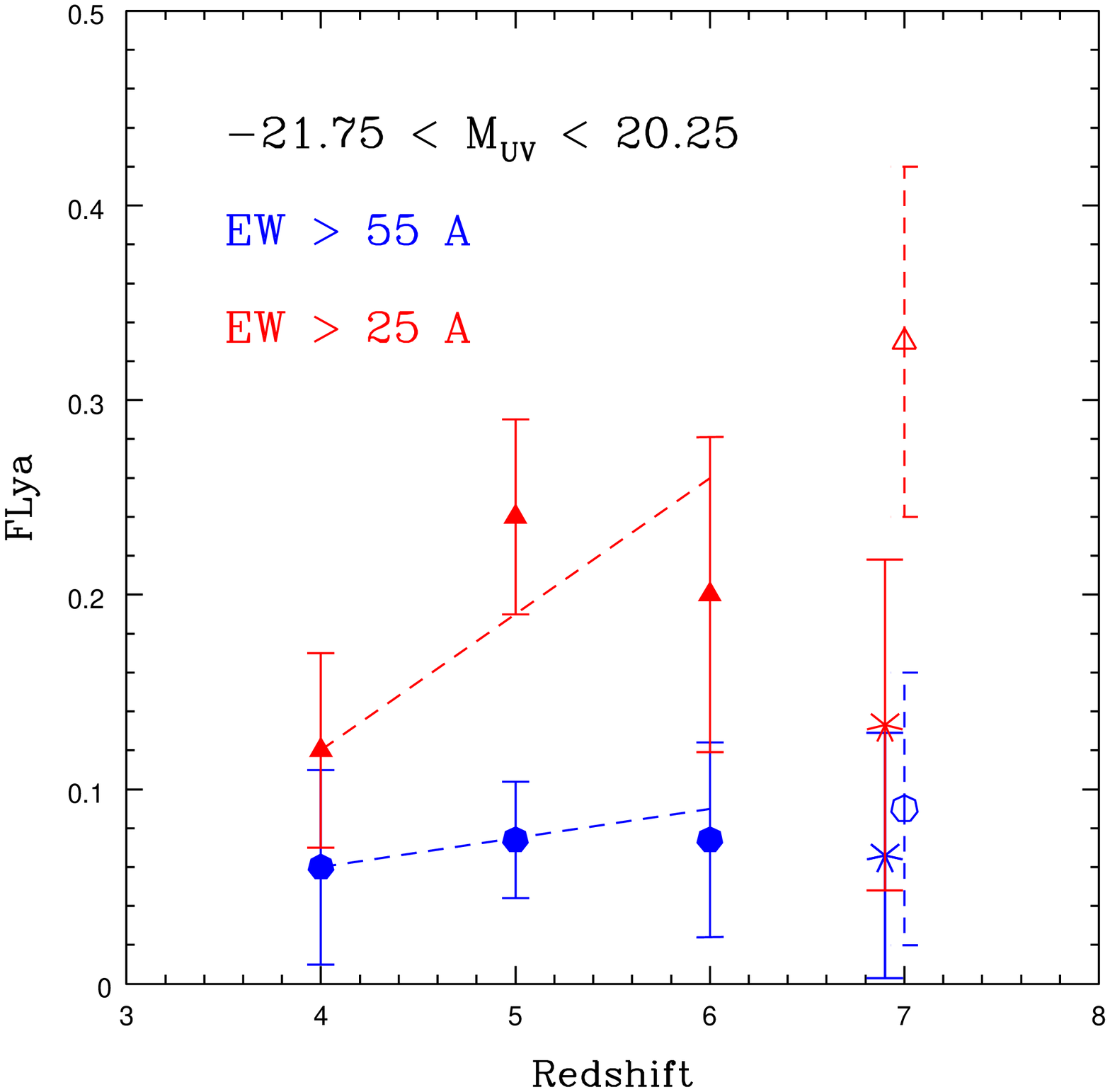}
\plotone{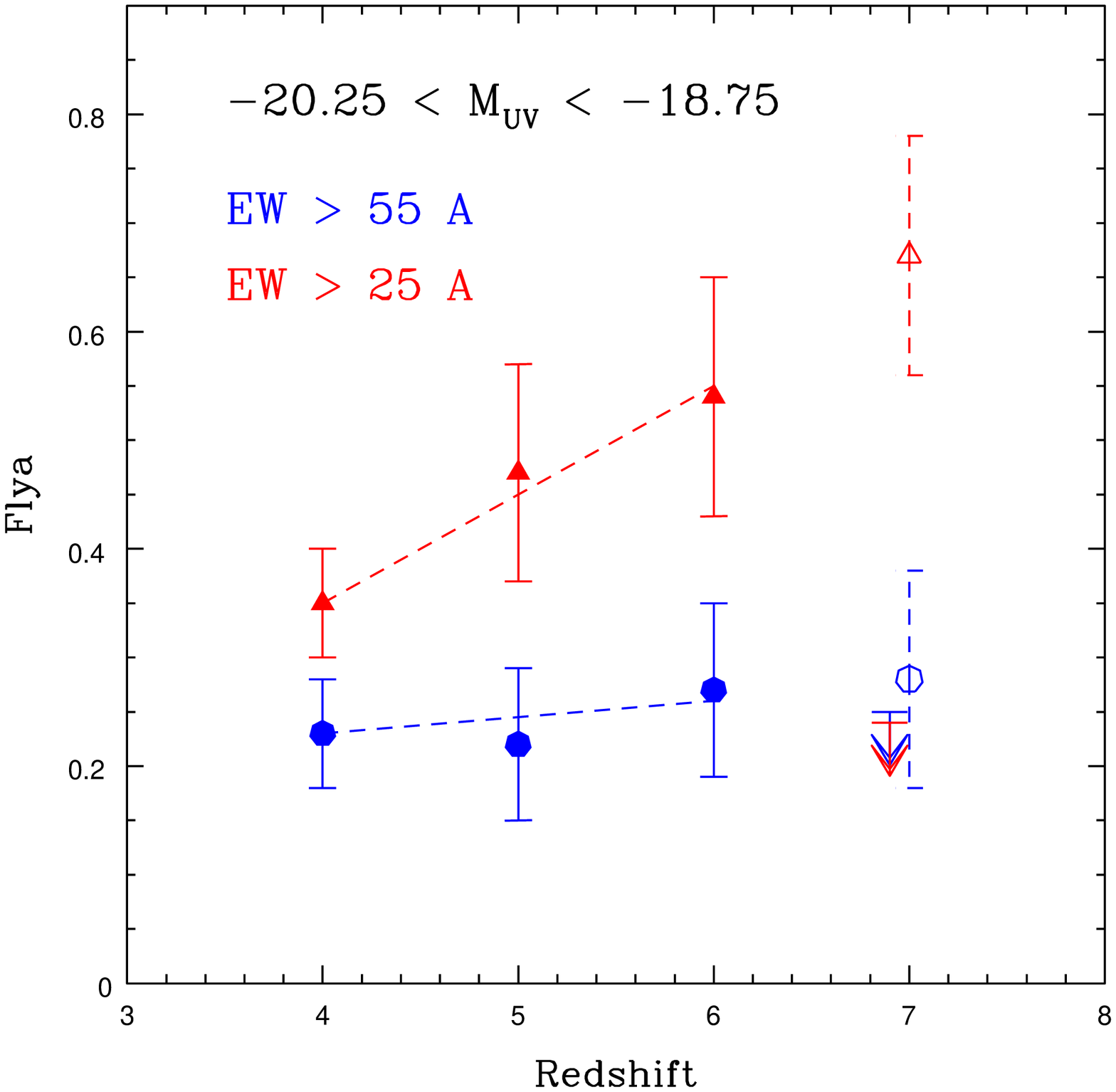}
\caption{{\it Upper panel}: The fraction of Lyman break galaxies with $-21.75< M_{UV}< -20.25$
 showing \lya emission at different redshifts and 
for different EW thresholds (25 \AA\ in red and 55 \AA\ in blue). The filled triangles and circles at  
$z\sim 4, 5$and 6 respectively are observations from S11; the open triangle/circle is 
their prediction for z$\sim 7$ on the basis of the fitted relation (dashed lines). 
The two asterisks are from this work. {\it Lower  panel} Same as upper panel for 
galaxies with absolute luminosity $-20.25< M_{UV}< -18.75$. The upper limits are from this work.}
\label{starkfig}
\end{figure}

Overall we have confirmed 5 galaxies at $6.7<z<7.1$ 
out of the initial 20 objects observed: of the remaining 15, object NTTDF--1917  is an interloper
(although its redshift is uncertain) and 14 are undetected.
The redshift distribution of the 5 confirmed galaxies is in 
agreement with the one expected from the color selection criteria and derived in C10a.
The obvious question to raise now is if this rate of 
spectroscopic confirmation and the general modest EWs of the \lya lines 
are in agreement with  what was expected on the basis of lower 
redshift surveys of LBGs.
\\
\begin{table}
\scriptsize
\begin{center}
\caption{Spectroscopic properties of confirmed  z$\sim 7 $ galaxies \label{tbl-2}}
\begin{tabular}{lllllll}
ID          &z            &  $L_{Ly\alpha}$            & EW   & S/N \\
            &             & erg s$^{-1}$               & \AA \\
\hline
BDF--3299   & 7.109        & 6.1$\times 10^{42}$ & 50 & 16 \\
BDF--521    & 7.008        & 7.1$\times 10^{42}$ & 64 & 18 \\
GOODS--1408 & 6.972        & 2.0$\times 10^{42}$ & 13 & 7 \\
NTTDF--6345 & 6.701        & 4.4$\times 10^{42}$ & 15 & 11 \\
NTTDF--474  & 6.623        & 1.7$\times 10^{42}$ & 16 & 7\\   
\hline
\end{tabular}
\end{center}
\end{table}
In the most recent and complete work on z$\sim 6$ galaxies 
\cite{Stark2011}(S11 from here on) found  a very  high  fraction of 
\lya  emitters  amongst Lyman break galaxies.
They assembled a sample of 74 spectroscopically observed 
z$\sim$6 galaxies by putting together their own Keck observations
 with all previous data available in the literature
(e.g. V09, Dow-Hygelund et al. 2007, Stanway et al. 2004, Bunker et al. 2003). 
With this, they were  able to confirm the redshift dependence 
of the fraction of line emission in LBGs,  more robustly 
and to higher redshift than was possible before 
(e.g. Stanway et al. 2007, Vanzella et al. 2009).
In particular they  found that 54\%  of faint ($M_{UV}> -20.25$) 
z $\sim$ 6 LBGs show 
relatively strong   (EW $>25$ \AA) emission and that a significant
 fraction (27\%) has $EW >55$ \AA.
For more luminous galaxies ($M_{UV}< -20.25$) the fractions are lower,
but still about 20\% of them are detectable as LAEs.
Moreover, by comparing these numbers to analogous samples of LBGs at
 z$\sim$4 and 5, they  
determined  that there is a strong increase of the Ly$\alpha$ emission fraction
with redshift. In particular the evolution  is very significant 
 for the intermediate EW  values with 
$d F_{Ly\alpha}^{25}/dz = 0.11\pm 0.04$, where $F_{Ly\alpha}^{25}$ is the fraction of 
LBGs with \lya\ stronger than 25\AA\ (in the original paper $\chi_{Ly\alpha}^{25}$)
 both for faint and bright galaxies. The increase is somewhat less pronounced for the brightest emitters with $EW>55$\AA\ \citep{Stark2011}.
\\
Their results suggest that, if the rise 
in the fraction of  LBGs that show prominent \lya emission observed 
over the epoch $4 < z < 6 $ continues to z$\sim 7$,  Ly$\alpha$ emission should be visible  in many  z$\sim$7 galaxies.
\\
It is therefore interesting  to make a comparison between 
the observed distribution of \lya emission in our sample and both
the z$\sim 6 $ observations by S11 and their predictions for z$\sim 7$.
For this comparison 
we will assume that all our photometrically selected 
dropouts are genuinely at z$\sim 7$, even if we failed to confirm them 
spectroscopically. The exception is the object NTTDF-1917
 that will be considered an interloper from here on
 and will be excluded from the sample.
This is the same assumption 
made by S11, who  found negligible contamination for luminous galaxies
$ (–22 < M_{UV} < –20) $
which make up also the large majority of our sample, 
 rising to only 10\%  for less luminous sources. 
\\
In Figure~\ref{starkfig} we plot the results of S11 together with their predictions for z=7
which were  derived by fitting  a linear relationship 
between the fraction of \lya emitter and 
redshift for $4 < z < 6 $, and then extrapolating  these trends to z$\sim$ 7.

To determine the fraction of galaxies in our sample 
that belong to each absolute  magnitude bin
 we proceeded in the following way:
for the galaxies with confirmed redshift, the absolute $M_{UV}$ was 
derived from the observed Y-band magnitude 
and the spectroscopic redshift taking into account 
the effects of Ly$\alpha$ forest absorption at that redshift 
and the \lya\ line flux  contribution.
They all have  $M_{UV}<-20.25$.
To assign an absolute magnitude $M_{UV}$ to the remaining 14 galaxies
 with no spectroscopic confirmation,
we proceeded in the following way. First, 
we randomly extracted  a redshift from the
N(z) distribution function corresponding to our initial z-dropout selection 
(see C10a, Figure 1). From this random  redshift and the observed 
Y-band magnitude  we derived $M_{UV}$.
We repeated this procedure 10000 times for each object and 
from the obtained $M_{UV}$ distribution, we  computed
the fraction of galaxies in the two bins 
$-21.75< M_{UV}< -20.25$ and $-20.25< M_{UV}< -18.75$.
In Figure~\ref{starkfig} (upper panel)  we plot as asterisks the fraction $f$ of galaxies 
in the high luminosity bin with EW $> 55$ \AA\ (F=0.066) 
and with $EW > 25$ \AA\ (F=0.132).
In the lower panel we plot as limits 
the corresponding fraction for the low luminosity galaxies,  
where we have no detections.
Our points  at z$\sim$6.9 for the galaxies with $EW > 25$ \AA\ are 
systematically   below the expectations 
derived from S11 for  z$\sim7$ and  are also lower than the observed fractions at z$\sim 6$.
We are ``missing'' a consistent number 
of objects that should show  \lya with intermediate EWs (between 25 \AA\ and 55 \AA), while the fraction of 
galaxies with very bright \lya\ emission (EW $> 55$ \AA) is consistent with the predictions. 
Note that this is not a due to a sensitivity limitation of our observations:
our EW detection limit
is well below $EW=25 \AA$  for  the whole spectral range probed 
and all galaxies with a broad band Y-magnitude brighter than 26.6, as shown in Figure 1 in F10
where we  plotted the 10$\sigma$ limit of the rest frame EW as a function of redshift achieved by our observations. As we mentioned in Section 2, the limiting flux of the BDF and NTT field spectra is  exactly the same as for the 
spectra of the GOODS-South field presented in F10.

As a final caveat we point out that we are comparing results at $z\sim 7$ obtained from homogeneous observations (in terms of set up, integration time and so  on) of a complete sample of galaxies, to  results that Stark et al. (2011)
derived from a compilation of very heterogeneous observations carried out on different samples. This could potentially introduce some bias although it is unclear in what direction.

\subsection{Assessing the significance of the EW evolution}
Given the small size of the sample we made several tests to 
estimate the significance of our result. We assume  that
the distribution of the \lya intensity in galaxies as a function of their rest-frame continuum magnitude $M_{UV}$ does not change significantly from z = 4-6 to z =7. 
We employ the same simulations developed in F10 and extensively described in that paper:
briefly, we assume that at $EW >0$\AA\
the EW distribution is represented by a Gaussian centered on $EW= 0$ \AA\ 
with an additional  constant tail up to 150 \AA, and at $EW <0$ by a constant level down to some EWmin value, and null below.
We take the width of the Gaussian and the two tails 
to reproduce the results of V09 and S10 at different rest-frame magnitudes, 
dividing the sample in two luminosity bins ($–20.5 > M_{UV}$ and
 $–20.5 < M_{UV} < –19.5$) and adjusting  the two tails to reproduce 
the fraction of galaxies 
with $EW>50$\AA\ given by S10 (for more details see F10).

We then compute the probability of detecting N \lya 
lines at a given S/N in our sample. 
For each object, we randomly extract a redshift from the C10a distribution, 
we compute the corresponding $M_{UV}$ from the observed 
Y-band magnitude and then we  randomly extract an EW from the 
corresponding distribution. If the EW is larger than the minimum detectable EW at the corresponding wavelength, we conclude that the object would be detected. 
We repeat this procedure 10000 times and in the end we compute the probability of having 3 detections with $S/N> 10$(or 5 detections with $S/N> 6$) as observed in our sample.

\begin{figure}
\epsscale{1.0}
\plotone{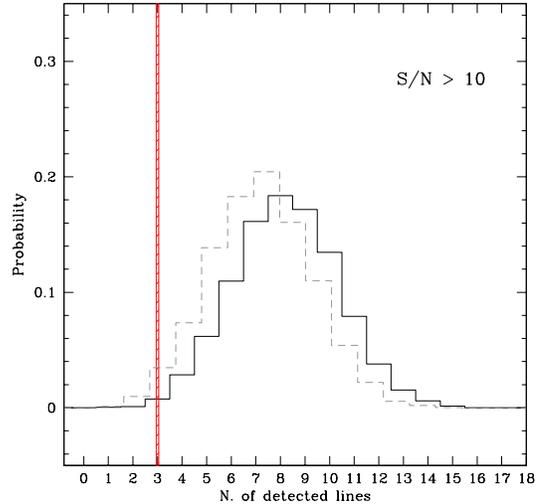}
\caption{The probability distribution of the number of expected detections 
with $S/N> 10$ in our sample: black represents the case assuming no interlopers and grey assuming that the faintest 3 objects are interlopers. The red vertical region indicates the observed results: in both cases the probability of obtaining our results is very low (1-3\%)}.
\label{probability}
\end{figure}

The results for $S/N>10$ are presented in Figure~\ref{probability}. If we assume
that all 14 undetected objects are bona-fide $z\sim 7$ galaxies 
the probability of finding  our result, given the input EW 
distribution, is extremely low, below 2\%.
Even if we consider  a consistent number of interlopers, and randomly
 eliminate  3 or 4 objects amongst the undetected ones,
 the probability is still as low as 3-6\%. The lower 
 probability values are found if we systematically
 eliminate the fainter objects 
from  our sample, since according  to the input distribution, they are expected
  show the brightest \lya\ lines. Vice-versa, eliminating all the bright objects, the probability of our results is slightly larger.

To reconcile our observations to the redshift 6 distribution, 
with a low but  acceptable probability (say $\sim $10\%), 
we have to remove $\sim$7 candidates from our sample, 
i.e. assume an interloper fraction of 50\%.
Note that if  we assume different EW distribution, such  as the one from 
Stanway et al. (2007), our result is even less likely, 
due to a much higher tail of large EW objects in their sample 
which we do not observe.

\begin{figure}
\epsscale{1.0}
\plotone{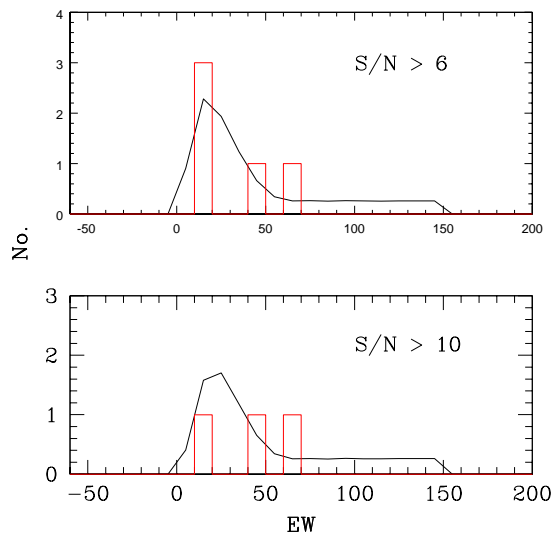}
\caption{The black lines represent the expected EW distribution 
of our sample assuming that the 19 objects are at z$\sim 7$ and follow 
the same distribution as the S10 i-dropouts. The red histogram are the 
 observed galaxies.
The top panel is considering objects detected with $S/N >6$ and the bottom panel for $S/N > 10$. The total number of objects expected is 9.7  in the first case (with 5 really observed) and 8.2 in the second (with 3 really observed)}.
\label{ewdist}
\end{figure}

Another way to visualize this result is by computing 
 the expected distribution of EWs derived  using the same simulations described above. 
The results are shown in Figure~\ref{ewdist} for  $S/N=6$  and  $S/N=10$.
In the first case we expect 9.7 detections and we find 5.
In the second case we expect 8.2 detection and we have only 3.
In both cases it is clear that the number of expected 
detections is significantly higher than what we obtain and  that, if the distribution of EW at z=7 was the same as  at z=6,  we should have detected more 
objects with intermediate EW (in the range 20-50 \AA)

We conclude that the discrepancy between the expected EW distribution of z-dropout galaxies and the observed one  is statistically significant. 
This  confirms the results found in F10, with a sample that is 
three times as large.

\section{Interpreting the drop of LAE fraction among LBGs:
an increasing neutral IGM ?}
Our  survey shows that there is  a significant change 
  in the fraction of \lya emitters amongst LBGs at $z> 6.5$: the trend for
the fraction of  \lya emission in LBGs  that is constantly 
increasing  from z$\sim$3 to z$\sim$6 
is stopped or most  probably  reversed at z$\sim$7.

Evidence for a decrease in the fraction of bright \lya  emission in galaxies 
at redshift  6.5 and beyond was first claimed by Kashikawa et al. (2006), who
noticed  an apparent deficit at the bright end of the \lya luminosity 
function of LAEs, compared to  that observed at $z=5.7$. 
At even earlier epochs (z$\sim 7$),
Ota et al. (2008) confirmed this  significant decrease 
in LAE density with increasing redshift, and interpreted it as due to
a combination of  galaxy evolution during these epochs and 
 attenuation of the \lya photons from LAEs by the neutral hydrogen 
possibly left at the last stage of cosmic reionization at z$\sim$6-7.
Subsequent searches for even higher redshift LAEs have 
all been unsuccessful so far (e.g. Clemens et al. submitted, Hibon et al. 2010).
However the picture, coming essentially from narrow band surveys for LAEs 
is still unclear, since other  authors claimed instead that there 
is hardly any evolution in the density of LAEs at redshift beyond 7 
(e.g. Tilvi et al. 2010) or that the evolution is only modest 
(Ouchi et al. 2010).

As already argued by other authors (e.g. F10, Ota et al. 2010), the drop in the 
frequency of bright \lya lines in galaxies could be due to 
different effects (or a combination of them):
(a) the LBG selection (z-dropouts) might suffer from an increasingly 
higher interloper fraction, such that a considerable number 
of our spectroscopic targets are not really at $z=7$ but at much lower z;
(b) there could be evolution in the intrinsic properties of the LBGs;
(c) an increasing neutral fraction of the IGM at z$>$ 6.5 could
suppress the \lya emission

We find it unlikely that a high fraction of interlopers 
is present in our sample for the following  reasons: 
our deep spectroscopic observations of i-dropouts (see  section 3.2)
show that the number of interlopers is actually rather modest, and that 
most of the objects that usually remain undetected in the  spectroscopic surveys
are actually at high redshift. We set a solid upper limit to the fraction 
of interlopers at $z=6$ of $<18\%$.
The selection of z-dropouts is done along the same lines as the i-dropouts. 
Great care was put in excluding any possible interloper (see C10a  for details)
such  that in the end  we  actually expect to have rejected a large
 number of real high redshift 
 galaxies from our sample, rather than include interlopers. For example for the GOODS-South field we estimated, though detailed simulations, that our strict selection criteria reject  about 30\% of real high-z galaxies (C10a), and the fraction is similar or even higher for the other fields.
Obviously we cannot exclude the possibility that a population of unusual objects are contaminating our sample, but there is no
reason why they should not appear also in the i-dropout spectroscopic sample.

Evolution in the  intrinsic properties of the galaxies, such as an increasing dust content,
could also provide an explanation
for the missing \lya. However as we go to earlier cosmic epochs we expect to find 
still younger objects, 
more metal and dust poor: in this case the production of \lya\ photons should be enhanced 
and we would expect to find an increasing fraction of \lya\ emitting galaxies (e.g. Shimasaku et al. 2006).  
Indeed  galaxies at very high redshift tend to show steep
UV slopes (e.g. V11, Bouwens et al. 2010), consistent with the hypothesis that these are relatively-dust
free systems. 

Alternatively, a very high escape fraction in high z-galaxies 
could  also reduce the intrinsic \lya~emission  (e.g. Dayal et al. 2008).  
The evolution of the escape fraction at high redshift in unknown at present (e.g. Haardt et al. 2011, Boutsia et al. 2011 and references therein). 
Furthermore for relatively high IGM neutral gas fraction, 
the \lya\  does not depend on $f_{esc}$  in a monotonic way (Dayal et al. 2008),
 since the effect of $f_{esc}$ is not only on the intrinsic 
\lya\ line but also on the Stromgren sphere size and hence on the visibility of the line itself 
(see also Santos 2004). For clarity we point out that  both 
Santos (2004) and Dayal et al. (2008)  are examples of 
homogeneous reionization models, in which the galaxy creates its own 
HII region. This introduces a direct relation between a galaxy's 
intrinsic luminosity and it's HII bubble size.
In more realistic inhomogeneous 
reionization models the HII bubble size 
is determined mostly by lower mass halos, which host less luminous galaxies. 
Hence the effects of a higher  $f_{esc}$ on the \lya\ visibility
is less certain.

Here we therefore attempt to interpret the apparent fast drop in the 
‘LAE fraction’ among LBGs
in terms of an evolving neutral hydrogen fraction. For this  we employ the 
models developed by \cite{Dijkstra2010}, which combine galactic outflow models 
with large-scale semi-numeric simulations of reionization  to 
quantify the probability distribution function of the fraction of \lya\
 photons transmitted through the 
intergalactic medium (IGM)  $\mathcal{T}_{\rm IGM}$.

Dijkstra et al. (2011) derive the $\mathcal{T}_{\rm IGM}$
by extracting  $\sim 10^4$ lines-of-sight (LOSs) centered on halos in the mass range, $10^{10}M_\odot < M_{halo} < 3 \times 10^{10}
M_\odot$. This choice of host halo masses was  motivated by the 
UV-derived star formation rate of few(2-4) $M_\odot yr^{-1}$ of the candidate
z=8.6 galaxy by Lenhert et al. (2010), which  corresponds to a halo mass of $>10^{10}M_\odot$ in the cosmological hydrodynamic simulations of Tran \& Cen (2007).  Our galaxies are similar or brighter than this object in UV therefore we assume they are hosted by halo of similar or higher masses.

In  Dijkstra et al. (2011), the observed EW-distribution function at z$\sim 6$ is modeled as 
an exponential function  which provides a good fit to the observed one at lower redshifts (e.g. Gronwall et al. 2007), with a 
scale-length that corresponds to the median value observed by Stark et al. (2010) 
of 50\AA. The IGM at redshift 6 is assumed 100 per cent
 transparent to \lya  photons emitted by galaxies: the EW probability distribution function  at $z= 7$  
is different from that at $z=6$ only because of evolution 
of the ionization state of the IGM.
The simulations assume no dust.

\begin{figure}
\epsscale{1.2}
\plotone{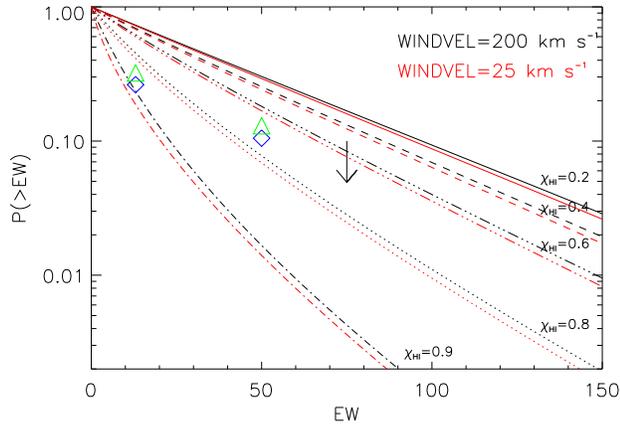}
\caption{The expected cumulative 
distribution function of Rest frame EWs for Z$\sim 7 $ LBGs, under the assumption that the observed LAE fraction at this redshift is different from z$\sim 6$only because of the IGM. The different lines correspond to a Universe that was respectively $\sim$ 0.2, 0.4, 0.6,0.8 and 0.9 neutral by volume (from top to bottom).
The black lines are for the wind model with $(N_{HI},v_{wind})=(10^{20} cm^{-2}, 200 km s^{-1})$ while the red lines are for  $(N_{HI},v_{wind})=(10^{20} cm^{-2},25 km s^{-1})$.
The blue diamonds (green triangles) are our results assuming 
that 0\%(20\%) or the undetected galaxies are interlopers.
The upper limit for EW=75\AA\ is from Stark et al. (2011).}
\label{dijkstra}
\end{figure}

This exercise is repeated for various fractions of neutral hydrogen (by volume), $\chi_{HI} =0.21,0.41,0.60,0.80,0.91$. The results are shown in 
Figure~\ref{dijkstra} 
for two models that differ only for the velocity of the outflowing  
winds assumed (25 and 200 km s$^{-1}$) and have the same column density $N_{HI}=10^{20} cm^{-2}$.
In case of higher velocity (black lines) 
the transmitted fraction of \lya\ radiation increases; the same is true
if the  column density is increased.

Our observations (blue diamonds) are consistent with  a neutral hydrogen fraction 
larger than  $\chi_{HI}=0.60$, even assuming that amongst 
the 14 undetections the fraction of  interlopers is as high as  20\%, 
(green triangles). This latter value was chosen assuming the percentage of interlopers at z$\sim 7$ is equal
to the upper limit we derived at z$\sim 6$.
In the same figure we also indicate the upper limit derived 
 by Stark et al. (2011) from their undetections at z=7, 
given their sensitivity to $EW>75$\AA.
Clearly there are many assumptions in the models, not least the fact that the 
Universe is considered completely reionized at z=6. 

For a comparison, Ota et al. obtained an estimate of  $\chi_{HI}^{z=7} \sim 0.32-0.62$ from  the lack of \lya emitters in the deep NB973 survey of
 the Subaru Deep Field  (Ota et al. 2008,2010) and using the Santos 2004 reionization models. 
Also, spectral analysis of z $\sim$6.3 and 6.7 GRBs imply that reionization 
is not yet complete at these epochs  with $\chi_{HI}^{z=6.3} \leq  0.17--0.6$
  (Totani et al. 2006;  but see also McQueen et al. 2008 for a different analysis) and 
$\chi_{HI}^{z=6.7} > 0.35$ (Greiner et al. 2009).
Finally the recent observations on the most distant QSO known at z=7.08 \citep{Mortlock2011}
 where analysed by  Bolton et al. (2011), who showed that the transmission profile is consistent
 with an IGM in the vicinity of the quasar with a volume averaged HI fraction of $\chi_{HI}>0.1$.

Overall these results, together with the high neutral fraction that we derive,
point to a rapid evolution  of the neutral fraction of hydrogen in the Universe between z$\sim 6$ and z$\sim 7$,  in required of the order $\Delta\chi_{HI} \sim 0.6$ in a short time ($\Delta z=1$).

\section{Summary and conclusions}
We have presented and discussed spectroscopic observations of a sample of 
20 z-dropout galaxies  selected from 
our deep and wide Hawki-I imaging program (C10a,C10b).
We confirm the redshifts of  
5 galaxies at $6.7 < z < 7.1$, through the presence of 
a \lya\ emission line. Only two of these galaxies have bright \lya\ and relatively high EWs.
The number of galaxies detected in our survey is considerably smaller than
what is expected from lower redshift surveys: the trend for an increasing fraction of \lya\ 
emission in LBGs that was found to hold from redshift 3 to 6 by previous studies (S11)
 is  halted and most probably reverted from z=6 to z=7.

If we assume that the intrinsic EW distribution of Ly$\alpha$
is that same at $z \sim 7$ as has been observed at $z \sim 6$, the discrepancy
between the predicted and measured EW distributions in our z-dropout sample
is statistically significant.
Our result would be marginally consistent with no evolution 
only if as many as half
of the unconfirmed galaxies were interlopers.  
The required interloper fraction would
be much larger than what we observe in our sample of i-dropouts, 
where interlopers are less than 
18\%, as well as what is commonly found in lower redshift samples (e.g. S10, V09).

These results extend and confirm the those of  F10, with a sample that is almost three times larger.

The  low detection rate, and the modest EW found can be explained in terms of  a rapid 
evolution of the neutral hydrogen fraction from z$\sim 6$ to z$\sim 7$.
Assuming that the Universe was completely reionized at z=6, our result point are consistent with a change of  neutral hydrogen of the order of $\Delta\chi_{HI} \sim 0.6$ in a relatively short time $\Delta z \sim 1$.
Indeed, there is no evidence at all that the Universe is 
completely ionized at $z=6$, e.g. Mesinger (2010) argues that, since reionization is expected to be highly inhomogeneous, much of the spectra pass through just the ionized component of the intergalactic medium (IGM) 
even for non-negligible values of neutral hydrogen fraction (volume mean)
 (see also McGreer et al. 2011) 
However  if we want to lower 
 substantially the $\Delta\chi_{HI}$ implied by our observations,  
then we need the Universe at $z=6$ to be  substantially neutral($>10$\%).

In any case, even if the models uncertainties are large, 
the absence of strong $Ly\alpha$ emission lines in our 
high  redshift galaxies  is very striking, and together with the very low number  of LAEs found in
all recent deep  narrow band surveys at z$\geq 7$ points to some drastic change that took place around 
that redshift. To better quantify the increase in the hydrogen neutral fraction at these epochs 
in the near future we will exploit the deep multi-wavelength CANDELS dataset (Grogin et al. 2011, Koekemoer et al. 2011) which will provide large numbers of candidate $z\sim7 $ galaxies
for further spectroscopic follow-up campaigns.

\acknowledgments
We acknowledge the support of ASI-INAF in the framework
of the program I/009/10/0 supporting Data Analysis in the field of Cosmology.








\end{document}